\documentclass[runningheads]{llncs}
\usepackage{graphicx}
\usepackage[ruled,vlined]{algorithm2e}

\usepackage{xcolor}
\usepackage{booktabs}
\usepackage{multirow}
\usepackage{float}

\usepackage{xcolor}
\usepackage{tablefootnote}
\definecolor{light-gray}{gray}{0.95}
\newcommand\Tstrut{\rule{0pt}{2.6ex}}         
\newcommand\Bstrut{\rule[-0.9ex]{0pt}{0pt}}   

\newcommand\justfootnote[1]{%
  \begingroup
  \renewcommand\thefootnote{}\footnote{#1}%
  \addtocounter{footnote}{-1}%
  \endgroup
}

\begin{document}
%
\title{Learned Indexing in Proteins: Extended Work on Substituting Complex Distance Calculations with Embedding and Clustering Techniques}

\titlerunning{Learned Indexing in Proteins}

\author{Jaroslav Olha\orcidID{0000-0003-1824-468X} \and
Terézia Slanin\'akov\'a\orcidID{0000-0003-0502-1145} \and
Martin Gendiar \and
Matej Antol \orcidID{0000-0002-1380-5647}\and
Vlastislav Dohnal\orcidID{0000-0001-7768-7435}}

\authorrunning{J. Olha et al.}

\institute{Faculty of Informatics, Masaryk University, Botanick\'a 68a, 602 00 Brno, Czech Republic
\email{\{olha, 492606, xslanin, dohnal\}@mail.muni.cz, antol@muni.cz}}
\maketitle              
\begin{abstract}
%
Despite the constant evolution of similarity searching research, it continues to face the same challenges stemming from the complexity of the data, such as the curse of dimensionality and computationally expensive distance functions. Various machine learning techniques have proven capable of replacing elaborate mathematical models with combinations of simple linear functions, often gaining speed and simplicity at the cost of formal guarantees of accuracy and correctness of querying. 

The authors explore the potential of this research trend by presenting a lightweight solution for the complex problem of 3D protein structure search. The solution consists of three steps -- (i) transformation of 3D protein structural information into very compact vectors, (ii) use of a probabilistic model to group these vectors and respond to queries by returning a given number of similar objects, and (iii) a final filtering step which applies basic vector distance functions to refine the result.

\justfootnote{\scriptsize{The publication of this paper and the follow-up research was supported by the ERDF "CyberSecurity, CyberCrime and Critical Information Infrastructures Center of Excellence" (No.CZ.02.1.01/0.0/0.0/16\_019/0000822). Part of this work was carried out with the support of ELIXIR CZ Research Infrastructure (ID LM2018131, MEYS CR). Computational resources were supplied by the project "e-Infrastruktura CZ" (e-INFRA CZ LM2018140 ) supported by the Ministry of Education, Youth and Sports of the Czech Republic.}}


\end{abstract}

\section{Introduction}

The methods and approaches developed by the similarity searching community are used by a wide range of scientific fields, both within computer science and beyond. While some applications require provable accuracy guarantees, verifiable algorithms, or support for complex similarity functions, many of the similarity searching problems emerging in various areas of research do not have such strict formal constraints.

Often, the representations of data or their similarity functions are inherently unreliable, so the idea of a perfect similarity search is unattainable regardless of the quality of the search method. Other times, the problems are too computationally complex, and instead of a perfect answer that would take too long to compute, a best-effort answer in a reasonable amount of time is preferable (perhaps to be refined later using more costly methods). Since the relevance of research on similarity searching is determined by its ability to solve current problems, it should strive to employ the widest possible array of problem-solving tools.


There are multiple similarity search indexes, falling under the umbrella of approximate searching, capable of adjusting to such use cases by lowering their accuracy thresholds or returning partial results~\cite{vadicamo2021generalizing}. However, in recent years, an entirely new approach has begun to gain traction -- along with most other areas of computer science, the area of data retrieval has started to incorporate various machine learning approaches. These approaches have the potential to not only expand the available toolbox of problem-solving instruments, but also to offer a new way of thinking about the problems themselves. Notably, in 2018, Kraska et al.~\cite{Kraska2018} suggested that all conventional index structures could be viewed as models of data distributions, implying that machine and deep learning models could be used in their place. Even though the idea was originally proposed and tested on structured data, this reframing of the problem has already inspired similar work in the realm of unstructured datasets~\cite{LMI2021}\cite{hunemorder2021}\cite{tian2022learned}.



To investigate the potential of these approaches further, we have chosen to examine the problem of 3D protein structure similarity search~\cite{LMIproteinsSISAP}. This is an important open problem in biochemical and medical research, which can be viewed as an instance of similarity searching in non-vector datasets, because similarity between a pair of protein structures is usually calculated using a series of non-trivial, computationally expensive operations. Additionally, the amount of 3D protein structure data is currently exploding due to a recent breakthrough in the field~\cite{alphafold2}, and the demand for versatile similarity searching approaches is likely to grow in the near future.

In this paper, we demonstrate that even a relatively complex interdisciplinary problem such as 3D protein structure retrieval can be tackled with fast and lightweight solutions. We present a simple pipeline where protein structures are first transformed into short vectors and used to train multiple partitioning and classification models -- these are linked together to form a learned index structure. The index then answers queries by returning several candidate leaf nodes, and filtering the objects stored therein using basic vector (similarity) functions\footnote{The entire functionality is publicly available as a search engine prototype at 
\url{https://disa.fi.muni.cz/demos/lmi-app-protein/}}.

This approach, while based on probabilities rather than mathematical guarantees, provides a reasonable quality of results at a fraction of the computational costs required by previous methods. In addition, its modularity allows us to change algorithms or their parameters for various trade-offs between complexity and accuracy, depending on the particular use case.

\section{Related work}

\textit{Learned indexing} was first introduced in~\cite{Kraska2018} with the core idea of learning a cumulative distribution function (CDF) to map a key to a record position. This proposition challenged the long-standing paradigm of building indexes solely with classic data structures such as B-trees and allowed for reduction in searching costs compared to the traditional methods. To allow for indexing of large data collections, the authors introduced \textit{Recursive model index (RMI)} -- a hierarchical tree index structure of simple interconnected machine learning models, each learning the mapping on a subset of the dataset. RMI is, however, limited to sorted datasets of structured data, and cannot accommodate multi-dimensional data.

The generalization of the learned indexing concept to spatial and low-dimen\-sional data was explored primarily by the \textit{Z-order model}~\cite{wang2019learned}, which makes use of the space-filling curve encoding to produce 1D representation of the original data, and \textit{ML-index}~\cite{davitkova2020ml} which achieves the same with the use of \textit{iDistance}~\cite{jagadish2005idistance}. \textit{RSMI}~\cite{qi2020effectively} introduced a recursive partitioning strategy, building on the Z-order model. Furthermore, \textit{LISA}~\cite{li2020lisa} and \textit{Flood}~\cite{nathan2020learning} both partition the data into grid cells based on data distribution, improving the range and kNN performance of prior approaches. These approaches were recently directly compared to (and surpassed by) \textit{LIMS}~\cite{tian2022learned}, which generalizes to metric spaces and establishes a new state-of-the-art performance on datasets of up to 64 dimensions.

Indexing solutions for approximate, rather than precise queries were explored by Chiu et al.~\cite{chiu2019learning} introducing a probability-based ranking model able to learn the neighborhood relationships within data, which can be integrated with existing \textit{product quantization (PQ)} methods. This method was tested on 1-billion datasets with approximately 100 dimensions and has been shown to boost the performance of traditional PQ methods.

Following the architectural design of RMI, we proposed the \textit{Learned metric index (LMI)}~\cite{LMI2021}, which can use a series of arbitrary machine learning models to solve the classification problem by learning a pre-defined partitioning scheme. This was later extended to a fully unsupervised (data-driven) version introduced in~\cite{slaninakova2021data}, which is utilized in this work.

Finally, the ideas of creating an indexing solution with machine learning have found their use in many different domains, e.g., in trajectory similarity search~\cite{ramadhan2022x} or information retrieval, using a single transformer model~\cite{tay2022transformer}.

\subsection*{Protein representation}

To enable computational approaches to the problem of uncovering functional properties of proteins, a great amount of research attention has been directed to creating representative (numerical) embeddings of protein structures. There are two distinct categories of embeddings based on the input data -- those that operate with \textit{sequences} and those working with \textit{3D structures}. Although they serve a similar purpose, these two categories are completely distinct in terms of the technical approaches they employ. Specifically, sequence embeddings use techniques such as hidden Markov models~\cite{koski2001hidden} or various natural language processing techniques~\cite{asgari2015continuous} to derive meaning from protein sequences, treating them as encoded sequences of characters -- this approach is not applicable to our research.




Embeddings representing protein 3D structures are generally less elaborate, since the information content is more robust to begin with. The most common encoding is a protein distance map, which produces a symmetric 2D matrix of pairwise distances between either atoms, groups of atoms or amino acid residues. This distance map can be transformed into a protein contact map, which is a binary image where each pixel indicates whether two residues are within a certain distance from one another or not. Contact maps have been used in conjunction with machine learning techniques for prediction of protein structure~\cite{cheng2007improved}; another studied problem is the reconstruction of 3D structure based on information in contact maps~\cite{vendruscolo1997recovery}. While these techniques are related to our own approach, we produce embeddings that are considerably more compact and reflective of our similarity searching use case, as will be shown in the following sections.

\section{Data domain}
We have chosen to test our approach on 3D protein structures for several reasons. First, while protein structure data is very widely used, and the study of this data is vital for almost every area of biochemical research, the issue of efficient search and comparison of protein structures is still unresolved to some extent, with many databases still relying on time-consuming brute-force linear search~\cite{mic2021similarity}. This data is also publicly available in a single database, called the Protein Data Bank (PDB), which is used by the majority of protein researchers and widely agreed upon as the standard. Even though this database is large enough to require efficient search methods, its size (\mbox{$\sim500,000$} structures as of 2022) still makes it possible to download the entire database and scan it exhaustively for ground truth answers if necessary.

Just as importantly, it is clear that the issue of efficient search within this data will only become more crucial and challenging in the next few years. Firstly, the common dataset of empirically solved protein structures continues to grow exponentially~\cite{burley2021rcsb} -- its current contents constitute a mere fraction of all the protein structures in nature, and the complex laboratory procedures needed to obtain these structures are being refined every year. Secondly, the computational prediction of protein structures from their sequences has recently seen rapid improvement with the release of AlphaFold 2 in 2021~\cite{alphafold2} -- this has already resulted in hundreds of thousands of new reliable 3D structures that are of great interest to researchers, with tens of millions more to be added in the coming months~\cite{varadi2022alphafold}.





Interestingly, the question of efficient protein structure search is not only important for storage and retrieval purposes -- for instance, researchers often discern the function of unknown proteins by comparing them to other, better-known proteins. Since the function of a protein is entirely dictated by its 3D structure, this is equivalent to searching a database for the most similar protein structures. However, the specific needs of this type of research can vary -- while some researchers are looking for extremely specific deviations among a group of very similar proteins (e.g., when studying mutations or conformational changes), others might be looking for much broader patterns of similarity between distant protein families.

Structurally, a protein chain consists of a linear sequence of interconnected building blocks called amino acids. Within the right biochemical environment, this linear sequence folds in on itself to form a complex 3D structure. 
Protein structures are sometimes cited as a typical example of complex unstructured data, since they cannot be meaningfully ordered according to any objective criteria (any search method needs to rely on pairwise similarity), and the similarity of two protein structures often cannot be determined by a single vector operation. Typically, protein molecule data are stored using the three-dimensional coordinates of each of their atoms, with the protein randomly oriented in space. In order to compare a pair of protein structures, they first need to be properly spatially aligned in terms of translation and rotation, and a subset of atoms must be selected for alignment. This typically involves gradual optimization of a spatial distance metric (such as the root-mean-square deviation of all the atom coordinates), which is a computationally expensive process that cannot be directly mapped to a simple vector operation.

One commonly-used measure of protein similarity is the $Q_{score}$~\cite{protein_distance}, which is calculated by dividing the number of aligned amino acids in both protein chains by the spatial deviation of this alignment and the total number of amino acids in both chains. Even though this measure is imperfect and not appropriate for all use cases, it is used in several prominent applications, including the PDB's own search engine.

Note that two identical structures have a $Q_{score}$ of 1, and completely different structures have a Q-score of 0: as a result, the score needs to be inverted in order to be used as a distance metric ($d(x,y) = 1 - Q_{score}(x,y)$). In the following sections, we will refer to this inverted value as $Q_{distance}$.

While there are a few types of protein structure embeddings that are invariant to the spatial alignment of the molecules (see the Related Work section), these were not developed for the purpose of fast data retrieval. As a result, they tend to be too detailed and cumbersome to serve as simple data descriptors. By contrast, the embedding presented in this work has been designed specifically to contain the optimal amount of information for efficient and accurate similarity searching, as will be shown in the following sections.

\section{Fast searching in proteins}


\begin{figure}[t]
    \centering
    \begin{center}
        \includegraphics[trim={.5cm .2cm 1.3cm .2cm}, clip, width=1\textwidth]{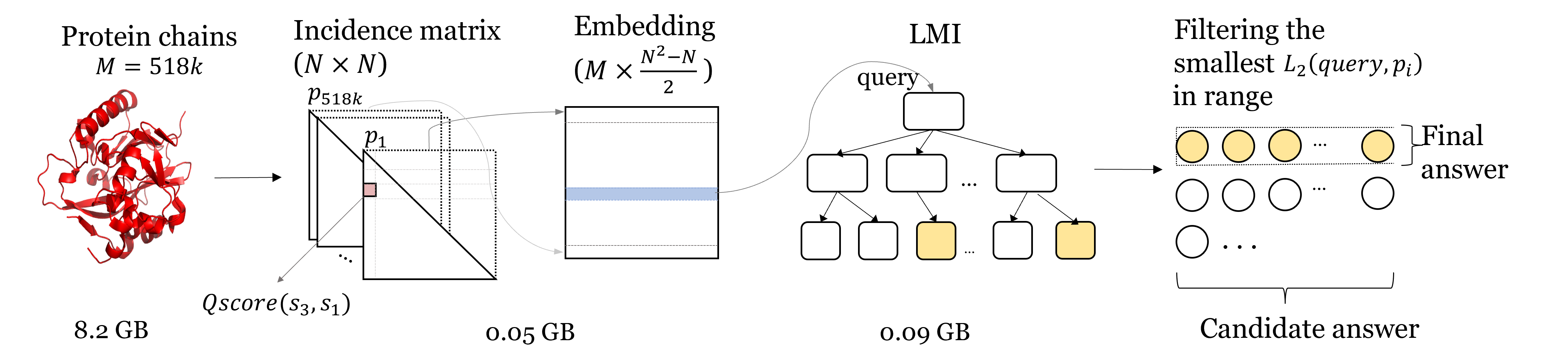}
    \end{center}
    \vspace{-5mm}
    \caption{A diagram of the proposed solution.}
    \label{fig:proteins_pipeline}
    \vspace{-2mm}
\end{figure}

We present a pipeline (visualized in Figure~\ref{fig:proteins_pipeline}) consisting of three separate components: (i) a simple embedding technique for protein data in the PDB format, (ii) the use of a machine-learning-based index -- Learned Metric Index (LMI) -- to locate a candidate set of similar protein structures, and (iii) fast filtering to produce the final query answer. 

The embedding we propose divides the protein sequence into $N$ consecutive sections -- the positions of the atoms within each section are averaged, and the section is subsequently treated as a single point in space. We then calculate distances between each pair of these sections, creating an incidence matrix. In this matrix, we prune all the values exceeding a cutoff, and normalize the rest. The matrix is symmetrical and all the diagonal values are 0. 
The half of this matrix (omitting the main diagonal) is then reduced into a single row in a $M \times (\frac{N^2-N}{2})$ matrix, where $M$ is the number of proteins in the entire dataset (see Figure~\ref{fig:proteins_pipeline}).

This produces a very compact embedding for all the proteins, and the entire dataset can be represented by a file that is up to two orders of magnitude smaller than the original database -- see Table~\ref{tab:descriptor_file_sizes}.

\begin{table}[t]
\centering
    \begin{tabular}{ |c|r|r|r| }
        \hline
        Embedding Size & File Size & Index Build Time (256-64) & Index Build Time (128-128) \Tstrut\Bstrut
        \\
        \hline
        $5\times5$    & $16$ MB   & 246s  & 184s  \Tstrut\Bstrut \\
        $10\times10$  & $51$ MB   & 350s  & 270s  \Tstrut\Bstrut \\
        $30\times30$  & $456$ MB  & 927s  & 655s  \Tstrut\Bstrut \\
        $50\times50$  & $1275$ MB & 2391s & 1814s \Tstrut\Bstrut \\
        \hline
    \end{tabular}
    \caption{File size of the protein dataset (518,576 protein chains) stored using protein embeddings, and build times of two different LMI architectures. Note that the size of the original database is 8.2 GB.}
    \label{tab:descriptor_file_sizes}
    \vspace{-5mm}
\end{table}

To reduce the search space to a small number of candidate protein structures, we used the Learned Metric Index (LMI), a tree index structure where each internal node is a learned model trained on a sub-section of data assigned to it by its parent node~\cite{LMI2021}. Specifically, we used the data-driven version of LMI, where the partitioning is determined in an unsupervised manner. We explored different architectural setups -- both in terms of the number of nodes at each level (index breadth), as well as the number of levels (index depth). As the learned models, we explored K-Means, Gaussian Mixture Models, and K-Means in combination with Logistic regression (see~\cite{slaninakova2021data} for details regarding the model setups). For the sake of compactness, in the experimental evaluation we only present the results achieved with the best-performing setup -- a two-level LMI structure with arity of 256 on level 1 and 64 on level 2 (i.e., 256 root descendants, each of them with 64 child nodes), with K-Means chosen as the partitioning algorithm. After LMI is built, we search within it using a query protein structure and return target candidate sets; the size of the candidate sets is determined by a pre-selected stop condition (for instance, a stop condition of 1\% of the dataset corresponds to $\sim5,000$ candidate answers per query).


In the final step, we filter the candidate set according to a particular distance function. In our experiments, we have examined filtering based on the Euclidean distance as well as the cosine distance of the vector embeddings, but the filtering step could theoretically be performed using any distance metric, or even the original $Q_{score}$ similarity of the full protein structures. The filtering step returns a subset of the candidate set based on the specified criteria (i.e., kNN or range).


\section{Experimental evaluation}
We evaluated our approach using range queries, with 512 randomly chosen protein chains from the dataset used as query objects. In order to compare our results against the ground truth, we needed to know the $Q_{distances}$ (based on $Q_{score}$) between the 512 protein chains and all the other chains in the database -- these distances were kindly provided by the researchers behind~\cite{mic2021similarity}, where the same 512 objects were used as the pivots for their search engine. The objects were chosen uniformly randomly with respect to protein chain length, which ensures that even very long proteins are represented among our queries (despite constituting a relatively small portion of the dataset).

\begin{figure}[t]
    \begin{center}
        \includegraphics[width=0.32\textwidth]{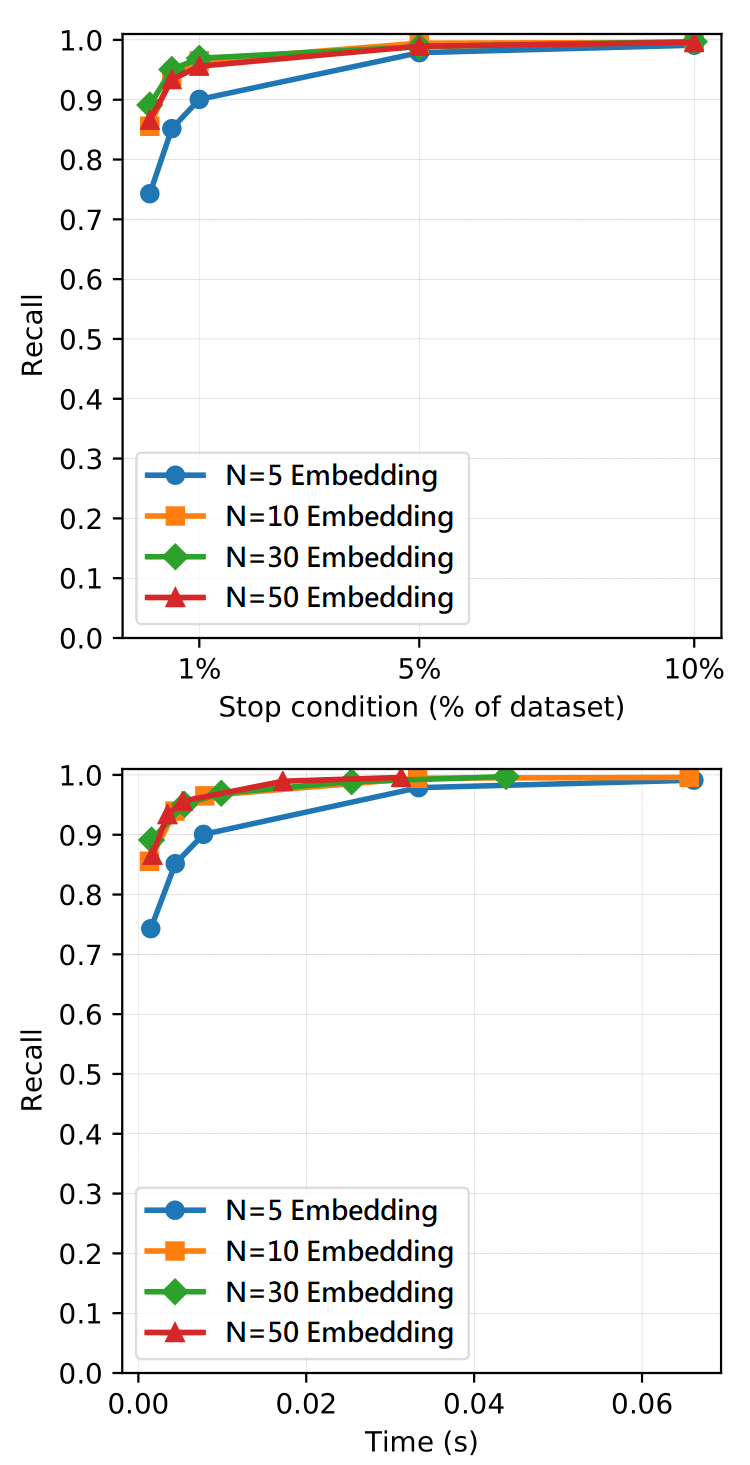}
        \includegraphics[width=0.32\textwidth]{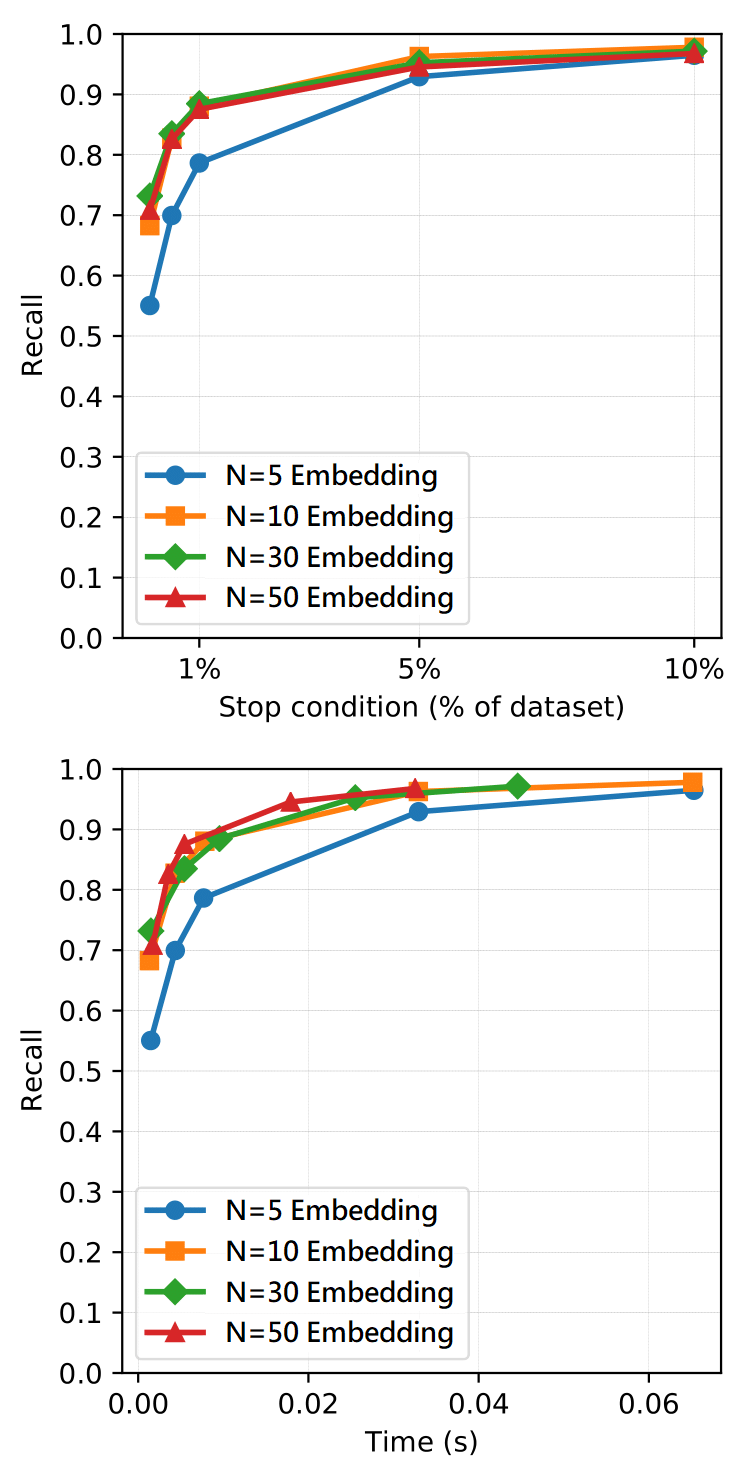}
        \includegraphics[width=0.32\textwidth]{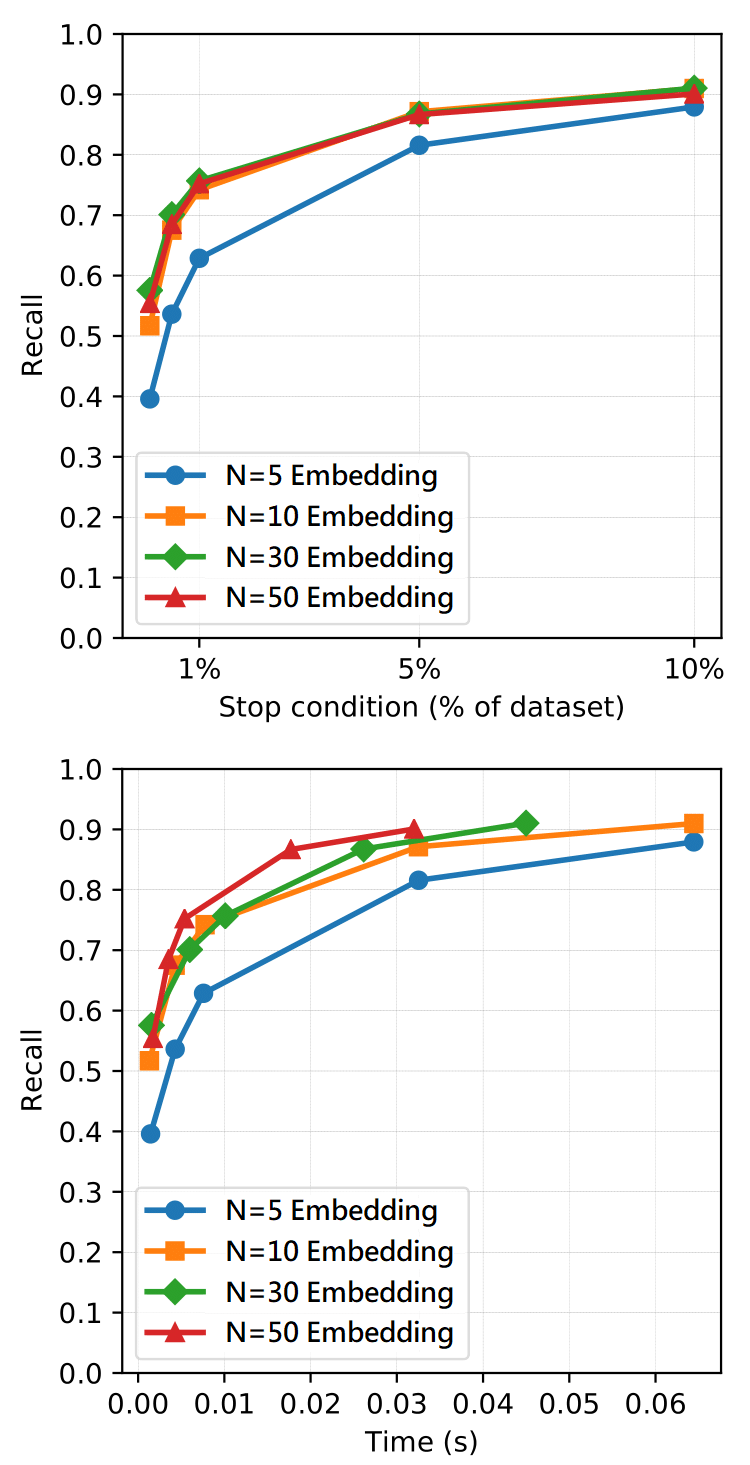}
    \end{center}
    \vspace{-5mm}
    \caption{Evaluation of range queries after LMI search and before filtering, using the K-Means method and a 256-64 LMI architecture: (left) Range=0.1, (middle) Range=0.3, (right) Range=0.5.}
    \label{fig:recall_265_64}
    \vspace{-2mm}
\end{figure}

We expected the performance of our method to deteriorate as the range of the queries expands, since a wider search range would require the method to correctly identify more objects which are less similar. To examine this effect, we have chosen three representative query ranges of 0.1, 0.3 and 0.5 -- in a real use case, the range would be chosen by a domain expert based on the particular use case. As a rule of thumb, a range of 0.1 represents a high degree of similarity, while a range of 0.5 represents low (but still biologically significant) similarity; the biological relevance of answers drops sharply beyond this range~\cite{mic2021similarity}.

First, we evaluated the performance of the LMI, before the filtering step. The recall shown in Figure \ref{fig:recall_265_64} pertains to the entire candidate set of objects (i.e. how much of the ground truth answer is contained in the 1\%/5\%/10\% of the dataset returned by the LMI for further filtering)\footnote{Precision is not evaluated in this step -- at this point in the pipeline, it is very low and not particularly relevant.}.


This figure presents us with two important pieces of information -- firstly, it is clear that LMI can reach very high recall even when trained on the smaller 10x10 embedding -- this makes the embedding a natural choice for further evaluation, since it is efficient while significantly reducing the memory and CPU costs of training compared with the larger embeddings. It can also be seen that, especially in the lower query ranges which are of most interest to us, the 1\% stop condition represents a sensible trade-off between recall and search time, returning relatively few candidate objects (\mbox{$\sim5,000$}) while minimizing the amount of false negatives.



\begin{figure*}[t]
  \centering
  \begin{minipage}[b]{0.35\textwidth}
  \includegraphics[width=1\textwidth]{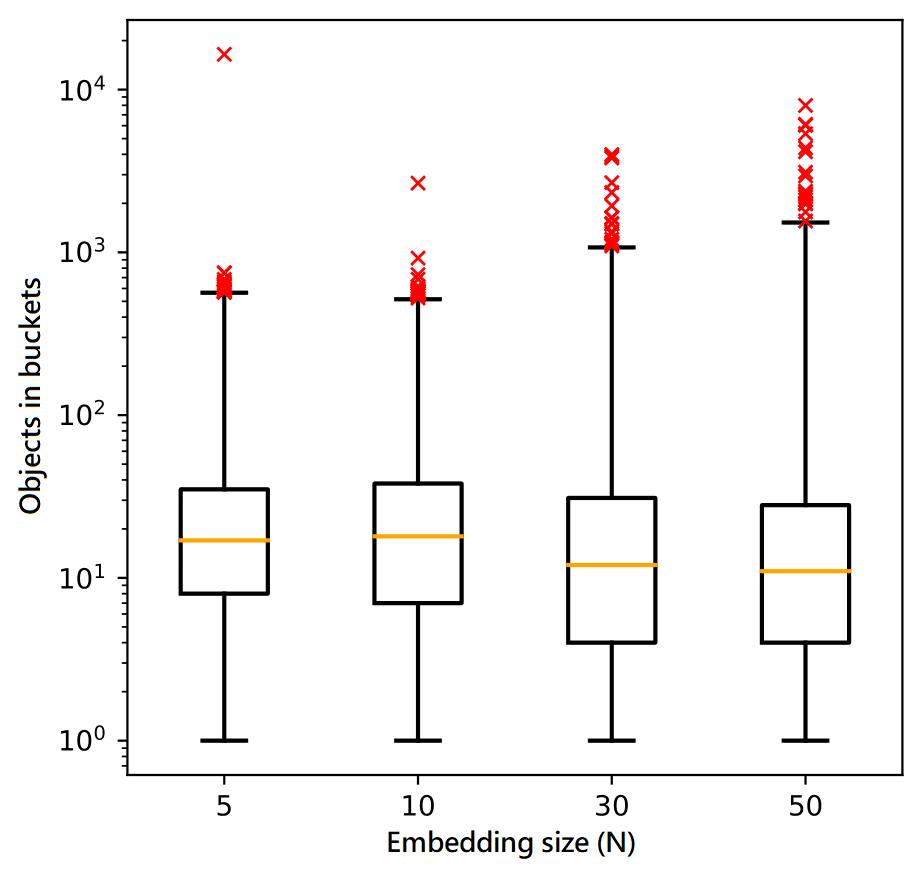}
    \caption{Distribution of objects in the LMI leaf nodes. (A~completely balanced structure would hold \mbox{$\sim30$} objects in each bucket).}
    \label{fig:bucket_size_distribution}
  \end{minipage}  
  \hspace{3mm}
  \begin{minipage}[b]{0.6\textwidth}
  \centering
  \includegraphics[trim={0 2cm 0 0cm},width=1\textwidth]{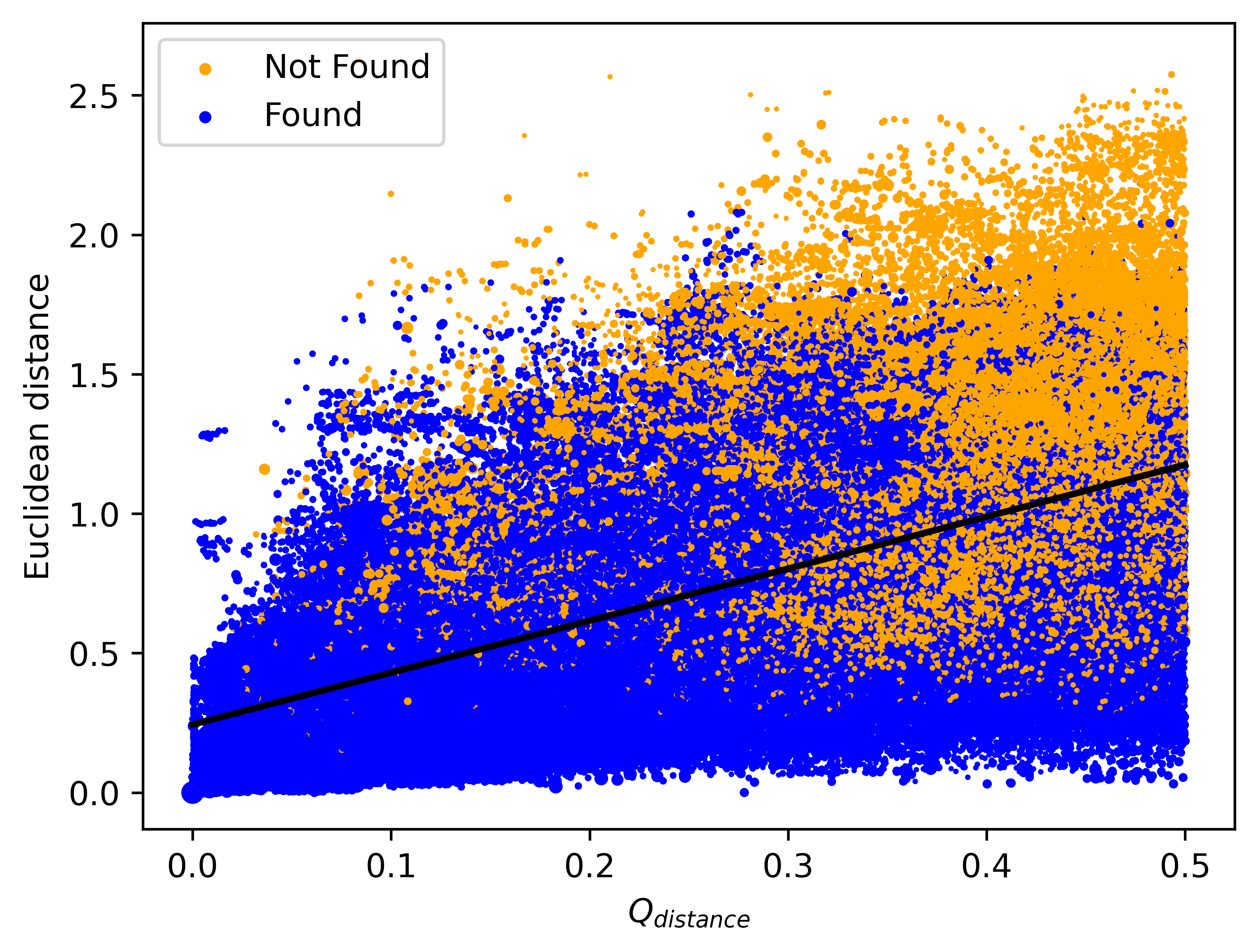}
    \caption{Correlation between $Q_{distance}$ and the Euclidean distance used for filtering: Blue points represent proteins returned by the LMI; orange points represent proteins which are present in the ground truth answer but were not returned by the LMI.}
    \label{fig:correlation}
  \end{minipage}  
  \vspace{-5mm}
\end{figure*}


In addition to recall, we also need to investigate the distribution of objects in the index, to ensure that the occupancy of leaf nodes (i.e. data buckets) is not overly imbalanced -- an extremely imbalanced index would achieve high recall, but it would also be more likely to return overly large candidate sets, which would be detrimental to the filtering step. The size distribution of the buckets is shown in Figure~\ref{fig:bucket_size_distribution}, and it confirms that 
the distribution is not overly skewed towards large buckets, even when the embeddings are quite small, as is the case with N=10. However, the embeddings can only be condensed up to a point -- the 5x5 embedding (which transforms each protein structure into a vector of only 10 values) causes a large portion of the objects to concentrate in a single bucket, as the LMI can no longer distinguish among groups of objects.

\begin{figure}[t]
    \begin{center}
        \includegraphics[width=0.32\textwidth]{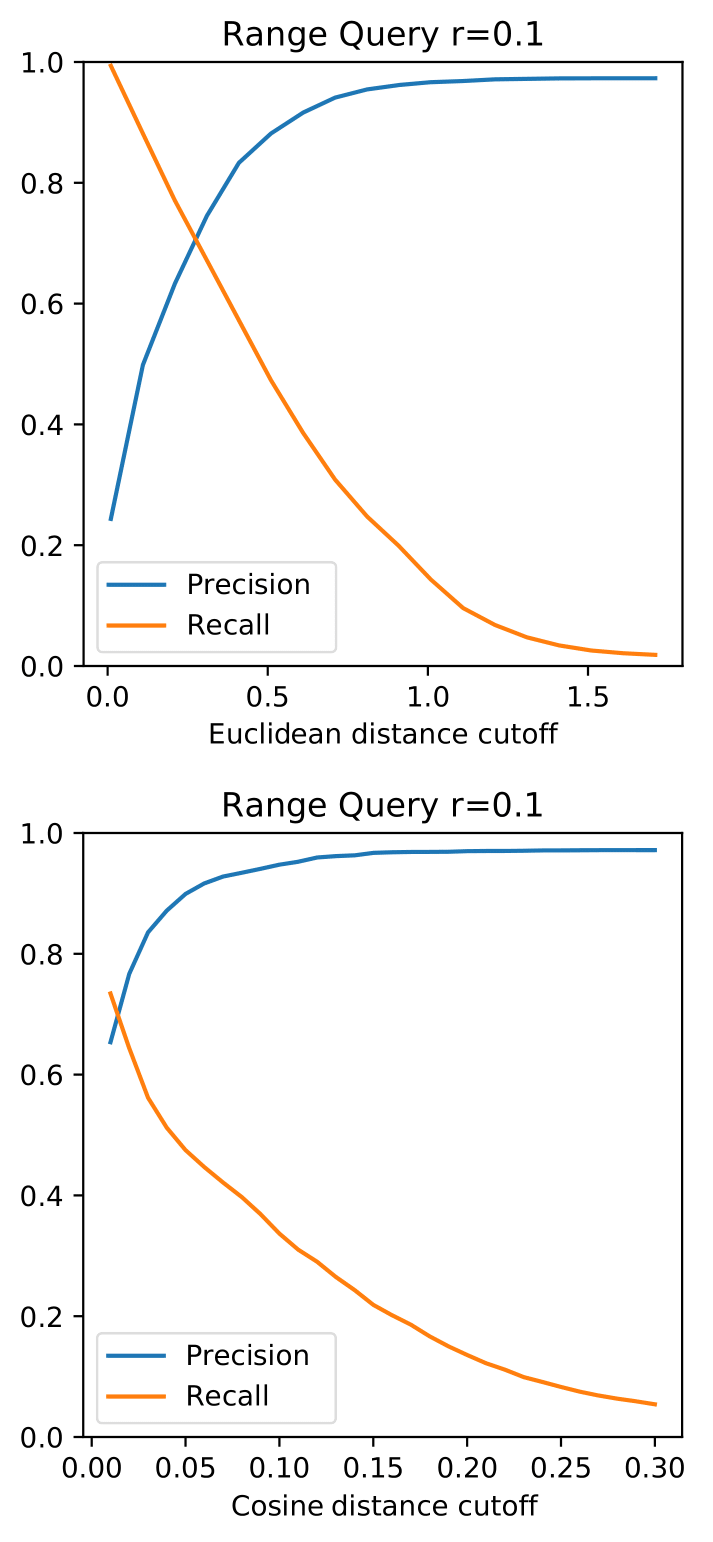}
        \includegraphics[width=0.32\textwidth]{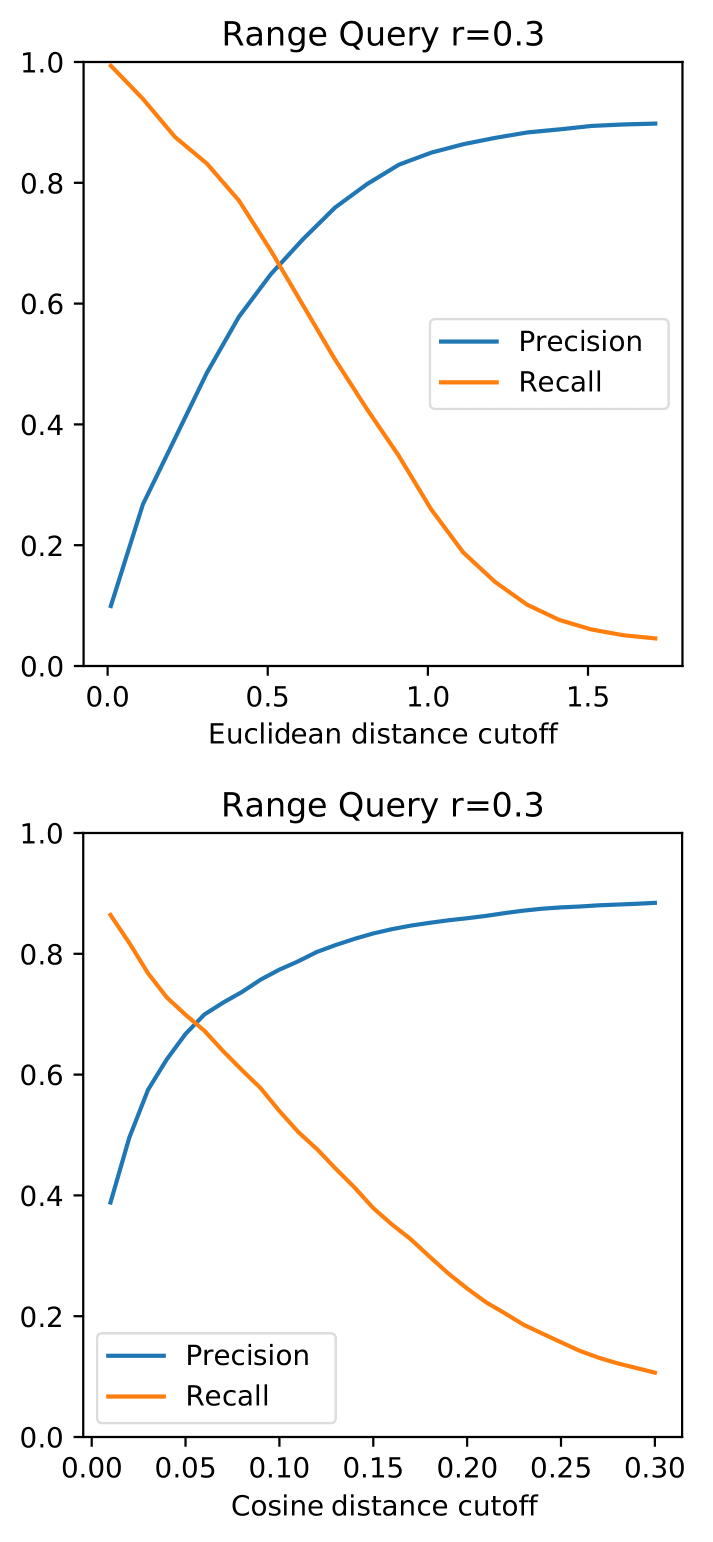}
        \includegraphics[width=0.32\textwidth]{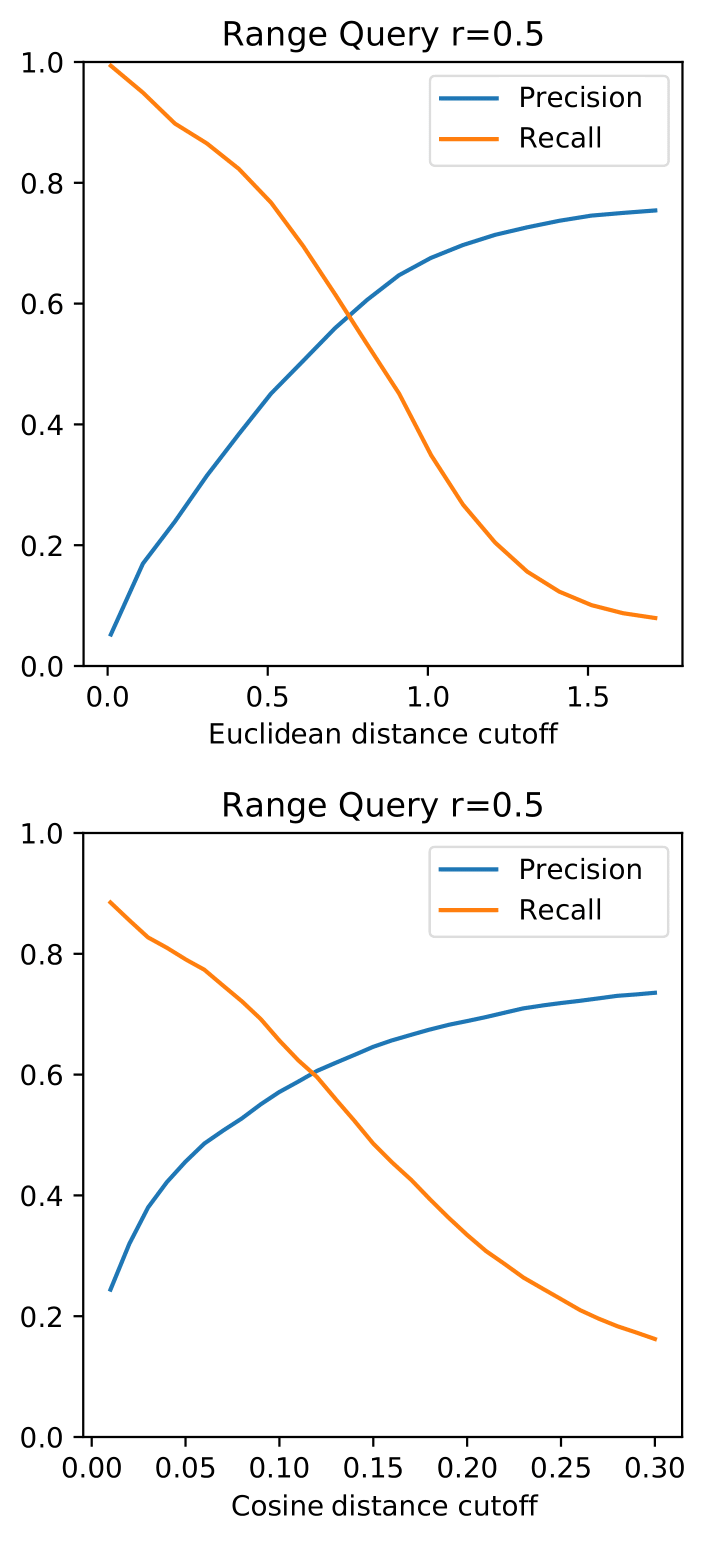}
    \end{center}
    \vspace{-5mm}
    \caption{Effects of filtering on the recall and precision of the candidate set of objects (relative to the ground truth answer). 
    }
    \label{fig:filtering}
    \vspace{-2mm}
\end{figure}

During the filtering step, recall naturally decreases over time (since the method occasionally filters out relevant answers), while precision should improve as the portion of relevant objects in the candidate set increases. This effect will be different depending on the distance metric used for filtering, as well as the dataset -- we show the effects for two distance metrics (Euclidean distance and cosine distance) in Figure~\ref{fig:filtering}. By analyzing the correlation between  $Q_{distance}$ and each of the two distance metrics, we have determined that the Euclidean distance is the better filtering function for this dataset.\footnote{Note that the actual $Q_{distance}$ function scales differently from the Euclidean distance used to filter answers (see Figure~\ref{fig:correlation}) -- this requires simple re-scaling (e.g., to find range=0.5 queries, we set the Euclidean distance cutoff at 0.75).}

Table~\ref{tab:pipeline_evaluation} shows the final results of the range queries with the best-performing configuration of parameters: embedding size N=10, the K-means clustering model, 256-64 LMI architecture, and filtering after the 1\% stop condition using the Euclidean distance metric. The results, especially in the lower query ranges, are very encouraging, although the filtering stage seems to introduce a surprisingly large amount of false negatives by filtering out parts of the correct answer. It is likely that the filtering metric we have chosen was slightly too na\"ive, and the filtering step could have benefited from a different distance function, or at least a different weighting of the vectors before calculating their Euclidean distance. In the future, this presents a natural point of focus to improve our results even further.

\begin{table}[t]
\centering
    \begin{tabular}{ |l|r|r|r| }
        \hline
        & Range 0.1 & Range 0.3 & Range 0.5 \\
        & \scriptsize{Mean \# of objects: 83} & \scriptsize{Mean \# of objects: 236} & \scriptsize{Mean \# of objects: 519} \\
        \hline
        LMI Recall & 0.973 (1.000) & 0.895 (0.999) & 0.755 (0.867) \Tstrut\Bstrut \\
        \hline
        Recall after filtering & 0.742 (0.878) & 0.649 (0.711) & 0.530 (0.637) \\
        \hline
        F1 after filtering & 0.712 (0.855) & 0.669 (0.766) & 0.592 (0.673) \Tstrut\Bstrut \\
        \hline
    \end{tabular}
    \caption{Overall evaluation of protein range queries: the average values, as well as the median values (in parentheses) are shown. 
    }
    \label{tab:pipeline_evaluation}
    \vspace{-5mm}
\end{table}



Figure~\ref{fig:correlation} further illustrates the strengths and weaknesses of our approach by showing the relationship between the $Q_{distance}$ metric and the Euclidean distances of vector embeddings. There is clear correlation between these two metrics, making a strong case for the simpler one, and LMI is much more successful at finding objects that are more similar to the query (left side of the graph) than it is at finding less similar objects (right side of the graph). It should be restated that even though we use the $Q_{distance}$ metric as the "ground truth", it is merely a subjective similarity metric, and the relevance of results that are close to the edge of the query range is debatable (i.e., an object with a $Q_{distance}$ of 0.499 is not necessarily more relevant than an object with a $Q_{distance}$ of 0.501).



\begin{figure}[t]
    \begin{center}
        \includegraphics[trim={.4cm 0 .5cm 0}, clip, width=.32\textwidth]{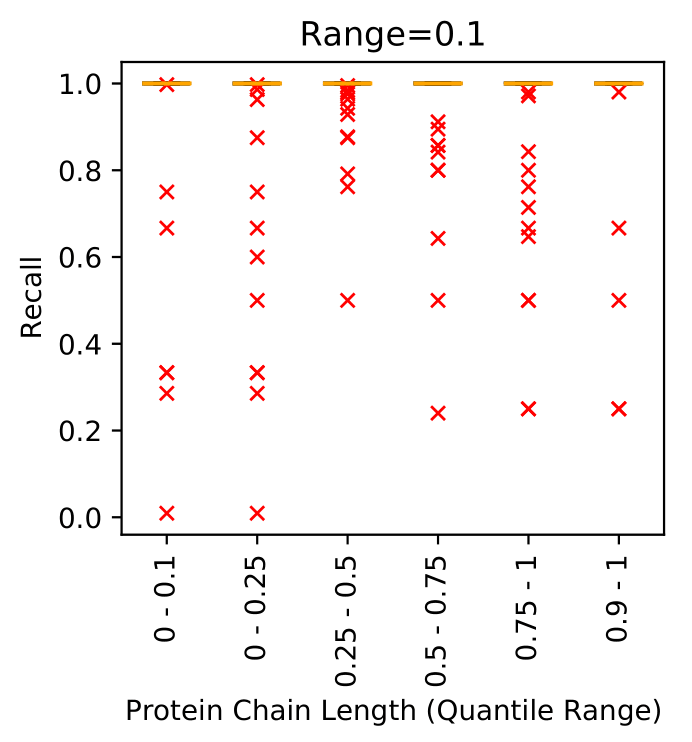}
        \includegraphics[trim={.4cm 0 .5cm 0}, clip, width=.32\textwidth]{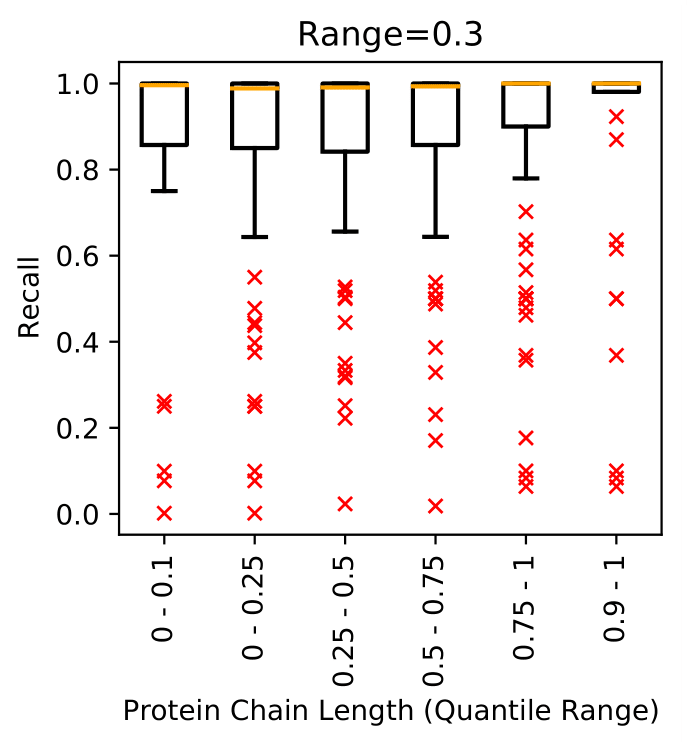}
        \includegraphics[trim={.4cm 0 .5cm 0}, clip, width=.32\textwidth]{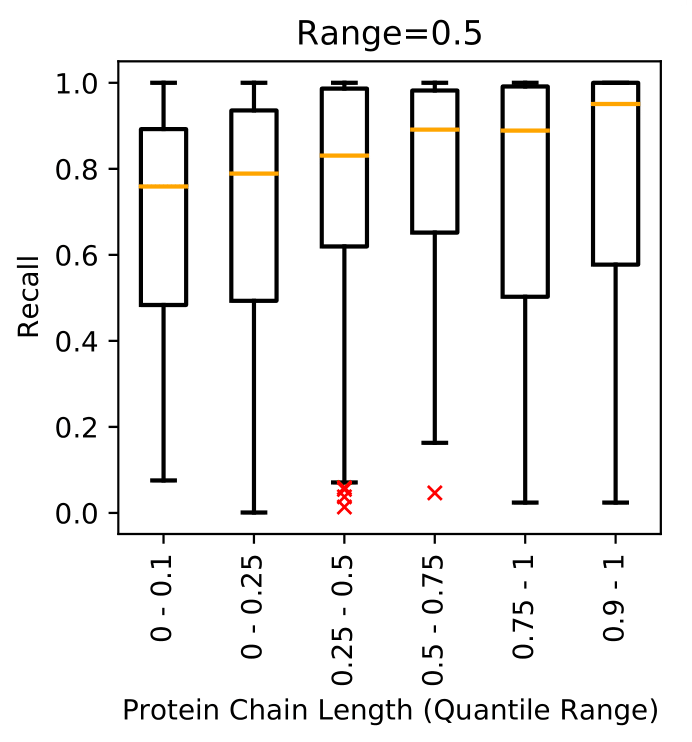}
    \end{center}
    \vspace{-5mm}
    \caption{Distribution of recall for different protein chain lengths (left to right: the shortest 10\% of chains, each quartile from shortest to longest, and the longest 10\% of chains).}
    \label{fig:protein_lengths_recall}
    \vspace{-2mm}
\end{figure}

Distance computations on long proteins pose a considerable challenge for similarity searching methods, typically requiring a disproportionate amount of computing time~\cite{mic2021similarity}. This does not apply to our embedding approach, which transforms all the proteins into fixed-length vectors using the same algorithm, regardless of their length. It would therefore intuitively follow that such an approach should achieve lower recall on the longer protein queries, since more information is lost in the embedding. However, in practice, that is not the case, as can be seen in Figure~\ref{fig:protein_lengths_recall}.

This is probably due to the simple fact that the distribution of chain lengths in protein databases is not uniform -- in fact, the long chains are much less numerous. As a result, losing a lot of information about the long chains is not a problem, since they are still relatively easy to find. This gives the fixed-length embedding approach a significant performance advantage -- in a dataset where there are relatively few long protein chains, it would be a waste of resources to calculate and store an excessive amount of data about them, the way many other methods do.

\begin{figure}[t]
    \begin{center}
        \includegraphics[trim={.4cm 0 .5cm 0}, clip, width=.32\textwidth]{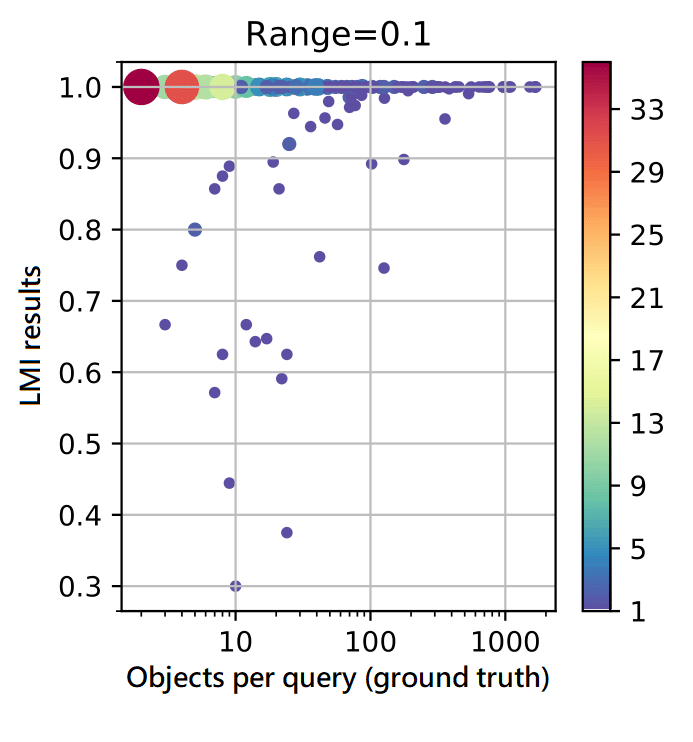}
        \includegraphics[trim={.4cm 0 .5cm 0}, clip, width=.32\textwidth]{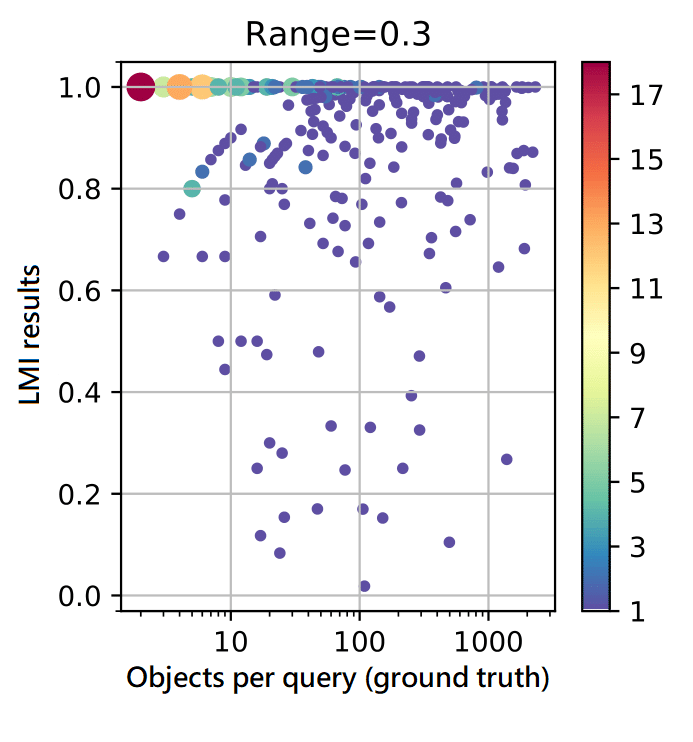}
        \includegraphics[trim={.4cm 0 .5cm 0}, clip, width=.32\textwidth]{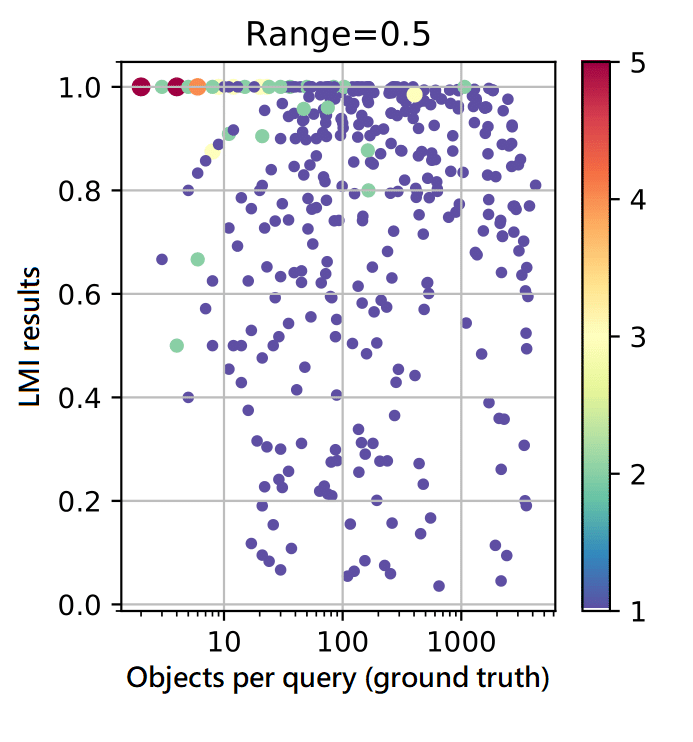}
    \end{center}
    \vspace{-5mm}
    \caption{The number of results returned by the LMI compared to the ground truth. The X-axis shows the absolute number of objects in a given query answer (according to the ground truth), while the Y-axis shows the portion of the objects that were found by LMI (i.e., the recall). Each blue point in the scatter plot represents one query, with the larger and more warmly-colored points representing multiple queries with identical graph coordinates.}
    \label{fig:heatmap_total_query_results}
    \vspace{-2mm}
\end{figure}

As has been mentioned in the previous section, in addition to variable chain lengths, the proteins also have a variable number of neighbors in any given range. This can, again, call into question the reliability of the results based on recall, since recall is a relative measure, and the size of the error will be different based on the size of the actual answer. For instance, if all range(0.1) query answers only consisted of a single easy-to-find object, evaluating recall on its own would give us a biased idea of the searching efficiency. Figure \ref{fig:heatmap_total_query_results} shows how many neighbors each protein structure has according to the ground truth, and how many of these neighbors have been found using our approach -- while there are significantly more proteins with fewer than 10 neighbors in the lower query ranges, these are not the majority, and the errors are distributed evenly relative to query answer size. 

Finally, to provide broader context for the pipeline's performance, we have evaluated it against a more conventional approach. While there is no analogous method for searching 3D protein structures using range queries, there is a similar, recently-published method for nearest neighbor retrieval in the same database of 3D protein structures~\cite{mic2021similarity}. This method uses a three-stage search engine which compares bit-strings in the Hamming space ("sketches") to approximate the distance of protein chains.

To allow for a relatively fair comparison of these two methods, we modified our search parameters to more closely match the search results presented in~\cite{mic2021similarity}. Specifically, since their similarity queries were mainly 30NN queries limited by the range 0.5, we have performed range queries in the same range (even though our method's performance is substandard in this range), and in the filtering step, we filtered out all objects beyond the 30 best-ranking answers. Since the sketch-based method was originally compared with the linear search of the PDB database, we present this benchmark as well -- naturally, this is always the slowest method by far, but it requires no index and always finds the exact answer.

All of these results can be found in Table~\ref{tab:state_of_the_art} -- while our method clearly does not match the high accuracy of the sketch-based method in this experimental setup, it is faster by at least an order of magnitude, occasionally exceeding 4 orders of magnitude since it does not suffer from an extreme "tail" of worst-case search times caused by evaluation of long proteins.

\begin{table}[t]
\centering
    \begin{tabular}{ |l|r|r|r| }
        \hline
        & LMI + Filtering \tablefootnote{Before the filtering step, the candidate set returned by LMI has a median recall of 1.0 and an average recall of 0.87; however, since the candidate set is much larger than the final answer ($\sim5,000$ objects), the accuracy is insignificant and has thus been omitted from the table.} & Sketch-based method & PDB Engine \Tstrut\Bstrut\\
        \hline
        Accuracy (median) & $0.660$ & $1.0$ & $1.0$ \Tstrut\Bstrut \\
        Accuracy (mean) & $0.626$ & $0.937$ & $1.0$ \Tstrut\Bstrut \\
        \hline
        Time (median) & $0.094$ s & $2.5$ s & $183$ s \Tstrut\Bstrut \\
        Time (max) & $0.145$ s & $6109$ s & $14321$ s  \Tstrut\Bstrut \\
        \hline
        Index size\tablefootnote{The size of the internal structure of the index, excluding raw protein data.} & $87$ MB & $178$ MB \tablefootnote{495,085 * (320b (sketches) + 1024b (sketches) + 4*6*64b (PPP-codes)) + 512 * 16kB (pivots)} & N/A  \Tstrut\Bstrut \\
        \hline\end{tabular}
    \caption{The accuracy, search times, and memory requirements of 30NN protein search queries with a maximum distance radius of 0.5.}
    \label{tab:state_of_the_art}
    \vspace{-5mm}
\end{table}

\section{Summary and Conclusions}
In an effort to investigate the potential of new data retrieval techniques in the field of similarity searching, we have developed and evaluated a novel approach to the problem of protein structure search, resulting in a short pipeline consisting of a concise vector embedding, learned indexing, and distance-based answer filtering. 
By successfully applying this approach on a well-established database of 3D protein structures, we have shown that even in a domain that may, at first, seem poorly suited to simple vector-based transformations, a surprising amount of information can be discerned by learned models.

One advantage of our modular approach is that every part of the pipeline can be evaluated separately, allowing experts to identify the weakest spots and alter them based on the current use case and dataset. The experiments presented in this paper serve as a good example -- after evaluating each part of our own pipeline, it is clear that we have chosen an overly simplistic filtering method for our data. In the future, we plan to investigate more sophisticated options for vector-based filtering, as well as a completely different approach to reducing the size of query answers. 

While it is difficult to compare our work with the state of the art (since there are no direct analogues to our method in the chosen data domain), we have made an effort to modify our method for the fairest possible comparison with a recent, more conventional similarity searching approach in the same domain. In this comparison, our solution, although coming up short in terms of accuracy, is consistently faster by multiple orders of magnitude, and maintains much lower memory requirements.

Our work aims to make a case for less conventional solutions to similarity searching problems -- ones that rely on learned approaches, rather than traditional indexing models, to discern the natural distribution of data. This is by no means an argument for universal adoption of machine-learning-based techniques in all similarity searching applications. The approach in this paper is problem-specific, and has several potential drawbacks that need to be weighed against its strengths. Nevertheless, we provide some insight as to how the trend of machine-learned pattern recognition, which has already caused minor and major revolutions in most fields of computer science, could be applied in practical solutions to current similarity searching problems.

Based on these results, we remain convinced that learned indexing approaches (such as the Learned Metric Index used in this work) will play an integral part in shaping the future of similarity searching. 


\bibliographystyle{splncs04}
\bibliography{main}
\end{document}